\documentclass[aps,pre,twocolumn,groupedaddress]{revtex4-2}
\usepackage{latexsym}
\usepackage{amsmath}
\usepackage{amsfonts}
\usepackage{amssymb}
\usepackage{amscd}
\usepackage{bbm}
\usepackage{bbold}
\usepackage{tikz}
\usepackage{tikz-cd}
\usetikzlibrary{shapes}
\usepackage{hyperref}
\usepackage[export]{adjustbox}
\usepackage[caption=false]{subfig}

\newcommand{\ket}[1]{|#1\rangle}
\newcommand{\bra}[1]{\langle#1|}
\newcommand{\bracket}[2]{\langle#1|#2\rangle}
\newcommand{\vev}[1]{\langle#1\rangle}

\renewcommand{\S}{\textsc{s}}
\newcommand{\I}{\textsc{i}}
\newcommand{\R}{\textsc{r}}

\newcommand{\tr}[1]{{\rm Tr} \left[ #1 \right]}

\begin{document}
	
\title{Complex information dynamics of epidemic spreading in low-dimensional networks}

\author{Wout Merbis}
\email[]{w.merbis@uva.nl}
\affiliation{Dutch Institute for Emergent Phenomena (DIEP), Institute for Theoretical Physics (ITFA), University of Amsterdam, 1090 GL Amsterdam, The Netherlands}

\author{Manlio de Domenico}
\email[]{manlio.dedomenico@unipd.it}
\affiliation{Department of Physics and Astronomy ``Galileo Galilei", University of Padua, Via F. Marzolo 8, 315126 Padova, Italy}
\affiliation{Istituto Nazionale di Fisica Nucleare, Sez. Padova, Italy}
 
\date{\today}

\begin{abstract}
	The statistical field theory of information dynamics on complex networks concerns the dynamical evolution of large classes of models of complex systems. Previous work has focused on networks where nodes carry an information field, which describes the internal state of the node, and its dynamical evolution. In this work, we propose a more general mathematical framework to model information dynamics on complex networks, where the internal node states are vector valued, thus allowing each node to carry multiple types of information.  This setup is relevant for many biophysical and socio-technological models of complex systems, ranging from viral dynamics on networks to models of opinion dynamics and social contagion. The full dynamics of these systems is described in the space of all possible network configurations, as opposed to a node-based perspective. Here, we illustrate the mathematical framework presented in the accompanying letter, while focusing on an exemplary application of epidemic spreading on a low-dimensional network.
\end{abstract}

\maketitle

\section{Introduction}

%Complex systems are characterized by emergent phenomena, which cannot be reduced to the individual constituents alone, but are the result of collective behavior due to their interaction among each other. Important examples ... \cite{}, Network interactions ... \cite{}, dynamical and out-of-equilibrium ... \cite{}

Collective behavior in complex systems can emerge as a results of the interplay between network structure, functionality and dynamics \cite{anderson1972more,crutchfield1994calculi,artime2022origin}.
Over the last few decades, much work has been devoted to understand these topics  \cite{watts1998collective,barabasi1999emergence,albert2002statistical,boccaletti2006complex,barrat2008dynamical,gao2016universal} and a general statistical field theory for information dynamics on networks has been under development \cite{ghavasieh2020statistical,ghavasieh2022statistical}, inspired by the parallels with quantum statistical physics and information theory.

The statistical field theory of information dynamics unifies a range of dynamical processes governing the evolution of information on networks, by defining an information field on top of each node. By means of a linearized equation for the dynamics of the field, a propagator $\hat{U}(t)$ quantifies the information flow between nodes in a complex network. The propagator $\hat{U}(t)$ can be used to define a density matrix $\hat{\rho(t)} = \hat{U}(t)/Z(t)$ with $Z(t) = \tr{\hat{U}(t)}$, whose von Neumann entropy is a measure of the functional diversity of the information streams \cite{de2016spectral}. Its maximal value indicates activation of all possible information streams, while it is minimized in the stationary distribution.

In many physically relevant situations, however, the linearized node-based approximation is inadequate due to significant correlations between nodes. Additionally, nodes may carry multiple types of information, reflecting the need for a theory of information dynamics where nodes have vector-valued information fields. 
A relevant example where these conditions are met is in network epidemiology \cite{hethcote2000mathematics,PastorSatorras2015,kiss2017mathematics}. Here, nodes can be in several different states, usually derived from compartments reflecting the disease progression in individuals. Dynamical rules then depend on the state of individual nodes, and are typically described by a set of coupled, \textit{non-linear} differential equations at the level of mean-field approximations \cite{pastor2001epidemic,may2001infection,van2008virus,van2011n}.

In an accompanying letter \cite{prldraft}, we advocate a generalization of the node-based information dynamics on complex networks to a more complete statistical description on the space of all possible network configurations. In this more general setup, the information dynamics can be understood in terms of information flow between different network configurations (microstates), as opposed to a node-based approach. The node-based dynamics emerges by taking relevant projections of the configuration space dynamics, together with a mean-field approximation. We demonstrate in \cite{prldraft} the relevance of the configuration-based approach by discussing two dynamical systems whose node-based dynamics is governed by the same propagator (the graph Laplacian), but the dynamics on the space of network configurations is wildly different: the voter model (VM) and the (symmetric) simple exclusion process (SEP) \cite{holley1975ergodic,liggett1985interacting,liggett1997stochastic}. 

Here, we illustrate the more general complex information dynamics at the hand of epidemic models on networks \cite{sahneh2013generalized,kiss2017mathematics,merbis2022logistic}. We focus on exact diagonalization, only feasible for small networks, to demonstrate the use of the general framework in the context of models of spreading phenomena on complex networks.  For practical applications, extensions to much large networks (and larger node state spaces) is necessary. In this case, powerful approximation methods from quantum-many body physics can be adapted to study the statistical mechanics of complex information dynamics, which we will discuss in the conclusions. We will leave the implementation of such techniques for network epidemiology to future work.

This paper is organized as follows: in section \ref{sec:infodynamics} we briefly discuss the general mathematical framework describing information dynamics on complex networks with vector-valued information fields, presented in \cite{prldraft}. We discuss applications of the general theory to much used models of epidemiology in section \ref{sec:applications}, including a derivation of the non-linear node-based dynamics, starting from the (linear) dynamical system at the level of network configurations. In section \eqref{sec:exactdiag} we discuss the exact diagonalization for network models on small sample networks. Finally, we present our conclusions in section \ref{sec:conclusion}.

\section{Information dynamics on networks}\label{sec:infodynamics}

Here we review the statistical field theory of complex information dynamics and present the main differences between a mathematical description based on the state of individual nodes and one based on the state of all configurations of the network as a whole. 

\subsection{Node-based dynamics}
The node-based description of complex information dynamics (see for instance \cite{ghavasieh2020statistical}) identifies the nodes of a network $i = 1, \ldots, N$ as basis vectors $\ket{i}$ of an $N$ dimensional vector space. A time-dependent information field $\ket{\phi(t)}$ can be defined on this vector space, such that the projection on node $i$, written as $\bracket{i}{\phi(t)}$, returns a scalar quantity which signifies the amount of information contained in node $i$ at time $t$. The dynamical evolution of $\ket{\phi(t)}$, after linearization is
\begin{equation}\label{nodebasedmastereqn}
	\partial_t \ket{\phi(t)} = F(t , A(t)) \ket{\phi(t)}\,,
\end{equation}
where $A(t)$ is the (possibly time-dependent) adjacency matrix of the network under consideration. Given a network model and its linearization, the solution to \eqref{nodebasedmastereqn} can be expressed in terms of a propagator $\hat{S}(t)$, such that $\bra{j} \hat{S}(t) \ket{i}$ represents the flow of information from source node $i$ to a target node $j$. 

This node-based description can describe a range of possible complex network dynamics, depending on the specification of $F(t, A(t))$, such as continuous-time diffusion, random walks, consensus, and synchronization \cite{de2017diffusion,masuda2017random,arenas2008synchronization,gomez2013diffusion}. However, there are also a number of limitations facing the node-based description. First off all, the dynamics \eqref{nodebasedmastereqn} is restricted to a linearized approximation of the true dynamics, which often is non-linear, especially when correlations between nodes are important. Second, the formulation allows only for a single type of information field for each node, whereas in some cases a more diverse set of different information types is needed to characterize the dynamics. For example, in contagion dynamics on networks, nodes take values in a set of compartments which characterize the state of the individual as susceptible, infected or otherwise. Dynamical transitions between node states may then depend on the specific state of neighboring nodes (i.e. whether neighbors are infected).

\subsection{Configuration space dynamics}
In our accompanying letter \cite{prldraft}, we have extended the complex information dynamics to characterize information flow not between specific nodes, but between all different configurations the network can take on. We suppose that the $N$ nodes of a network are random variables $X_i$ with $0 < i \leq N$, which may take $d$ distinct values in a discrete and finite set of states $\Sigma$, such that $d = \text{dim}(\Sigma)$. 
The state of the entire network $X = X_1X_2\ldots X_N$ is fixed by determining the states $X_i = s_i \in \Sigma$ for all $i$. 
%As each node can take on $d$ values, there are $d^N$ possible network states. Therefore, we can represent the probability of $X$ taking values $s_1 s_2 \ldots s_N$ as a $d^N$ dimensional vector, whose basis is conveniently represented using the tensor product basis
All $d^N$ possible network configurations $X$ span a vector space, which can be decomposed into a sum over tensor products of $N$ $d$-dimensional node vectors $\ket{s_i}$. The probability vector, whose elements correspond to the probability of encountering the system in a certain microscopic configuration $X$ at time $t$, can then be written as:
\begin{widetext}
\begin{equation}\label{manybodyvec}
P\left(X = \prod_{i=1}^N s_i\right) = \ket{\Phi} =  \sum_{s_1, \ldots, s_N \in \Sigma} P(X_1 = s_1; \ldots;  X_N = s_N) \ket{s_1} \otimes \ldots \otimes \ket{s_N}
\end{equation}
\end{widetext}
As outlined in \cite{prldraft}, for large classes of network models the configuration state vector  $ \ket{\Phi(t)}$  evolves in time by a master equation:
\begin{equation}\label{mastereqn}
	\partial_t \ket{\Phi(t)} = \hat W(A(t)) \ket{\Phi(t)} \,,
\end{equation}
where the Markov generator $\hat{W}(A(t))$ can be decomposed into a sum over (at most) bilocal operators
\begin{equation}\label{genericW}
	\hat W(A(t)) = \sum_{\lambda} r_\lambda  \sum_{i,j = 1}^N A^{ij}(t) \hat a^i_{\lambda} \, \hat b^j_{\lambda} + \sum_{\gamma} \sum_{i = 1}^N r_\gamma \hat c_{\gamma}^i
\end{equation}
Here $\lambda$ labels the bilinear (nearest-neighbor) transitions of the model (each occurring with rate $r_\lambda$) and $\gamma$ labels the local operators responsible for node transitions independent of the state of a nodes direct neighbors (and occurring with rate $r_\gamma$). The local operators $\hat a_\lambda, \hat b_\lambda, \hat c_\gamma$ appearing above are $d$-dimensional matrices, which only act on the local vector space of node $i$, leaving all other sites invariant. Their representation as a $d^N$ dimensional matrix is obtained by inserting at the node position $i$ the local operator $\hat a_\lambda, \hat b_\lambda$ or $\hat c_\gamma$ into a tensor product with $N-1$ identity matrices. 

The restriction to models generated by \eqref{genericW} implies practically that we are considering Markovian dynamics on the space of network configurations, where single node transitions are induced by the state of the local node environment (defined as the nodes own state and that of its direct neighbors). The setup can be generalized straightforwardly to higher-order interactions by including terms with three or more local operators acting simultaneously. In such case, one should additionally specify a hypergraph, which contains information on which simplexes are present in the model.

As a consequence of the conservation and non-negativity of probability, the Markov generator $\hat{W}(A(t))$ is an infinitesimal stochastic operators. This implies that its columns sum to zero and all off-diagonal entries are non-negative. As this property must hold for each term in \eqref{genericW}, the local operators $\hat c_\gamma$ and the tensor product $\hat a_\lambda \otimes \hat b_\lambda$ should also be infinitesimally stochastic. An explicit basis for these operators can always be constructed from the matrices
\begin{equation}
\hat q_{k \to l} = \ket{l}\bra{k} - \ket{k}\bra{k} \,, \qquad k,l \in \Sigma
\end{equation}
For the bilinear interactions, a similar basis is constructed by taking $k,l \in \Sigma \otimes \Sigma$. In this case, it might not always be possible to decompose the $d^2$ dimensional infinitesimal stochastic matrix into a direct product form $\hat{a}_\lambda \otimes \hat{b}_\lambda$. However, one can (by Schmidt decomposition) always write such a matrix as the sum over direct product matrices and use those to construct the bilinear interaction term in \eqref{genericW}.

\subsection{Spectral entropy of the configuration space model}
In analogy to the spectral entropy of the node-based dynamics \cite{de2016spectral}, in \cite{prldraft} we define the spectral entropy of the dynamical system on the configuration space as the von Neumann entropy of a density matrix $\hat{\rho}(t)$
\begin{equation}\label{spectralS}
	\mathcal{S}(t) = - {\rm Tr} \left[ \hat\rho(t) \log \hat\rho(t) \right] \,,
\end{equation}
where the density matrix $\hat{\rho}(t)$ is constructed as
\begin{align}\label{rhodef}
	\hat U(t) & = \exp\left( \hat W(A) t\right)\,, \\
	Z(t) & = {\rm Tr}\left[ \hat U (t) \right] \,, \\
	\hat\rho(t) & = \hat{U}(t) /Z(t)\,.
\end{align}
These terms have the following interpretations in terms of the dynamical process. $\hat{U}(t)$ is a right-stochastic matrix, whose entries $\bra{Y}\hat{U}(t)\ket{X}$ correspond to the conditional probabilities of finding the network in the configuration $Y$ at time $t$, given that it was in configuration $X$ at $t=0$. As such, $Z(t)$ signifies the sum over all return probabilities (the probability of finding the system in microstate $X$ at time $t$, given that it started in $X$ at $t=0$) or, the total amount of information field trapped in the microstates. The density matrix $\hat{\rho}(t)$ combines these two objects into a single normalized matrix, in analogy with the density matrix formulation of quantum mechanics. Note, however, that this density matrix will not necessarily be Hermitian; only when the stochastic process satisfies local detailed balance, a similarity transformation exist between $\hat{W}(A)$ and a Hermitian matrix. In such cases, a genuinely quantum mechanical density matrix can be defined for the process in terms of a positive semi-definite operator. In any case, $\hat{\rho}(t)$ as defined here is still an interesting object to study, as it encodes both the amount of trapped field (in $Z(t)$) and the diversity of information flows (through $\hat{U}(t)$).

The spectral entropy \eqref{spectralS} gives a measure of the diversity of information streams in the system. It is upper bounded by $\mathcal{S}(t) \leq N \log d$, when $\hat \rho(t) = \frac{1}{d^N} \mathbb{1}_{d^N \times d^N}$. In this situation, each microstate has uniform `trapped field', such that all information streams are activated. At late times, the spectral entropy gives a logarithmic measure of the number of steady (or absorbing) states in the system, as all information streams eventually end in a steady or absorbing state. 

At intermediate times, the spectral entropy can be written explicitly as
\begin{equation}\label{SvNexplicit}
\mathcal{S}(t) = - \tr{\hat{W}\hat{\rho}(t)} t + \log Z(t)\,.
\end{equation}
This expression consists out of two terms, both of which are manifestly positive. The first term can be rewritten in terms of the rate of change of the trapped field:
\begin{equation}
- \tr{\hat{W}\hat{\rho}(t)} t = - Z(t)^{-1} \partial_t Z(t)
\end{equation}
As such, this term quantifies the flow (or dissipation) of information out of the trapped field $Z(t)$. The second term in \eqref{SvNexplicit} is a logarithmic measure of the amount of trapped field $Z(t)$. This decomposition of spectral entropy invites an analogy to thermodynamics, where one could interpret the log of the trapped field with internal energy and the dissipation of trapped field as heat flow.

\subsection{Recovering the node-based dynamics from the configuration space model}\label{sec:nodebasedobs}

The configuration space propagator $\hat{U}(t)$ contains all (conditional) probabilities of the dynamical evolution of the system. As mentioned before, its elements directly correspond to the conditional probabilities of transitioning between certain microscopic configurations at time $t$. From this, one can recover the node-based dynamical evolution of the system, by making suitable projections on a collection or distribution over microstates. 

For instance, the probability of finding the node $i$ in state $s \in \Sigma$ is given by projecting the space of all configurations to only those in which node $i$ is in the state $s$. We define the (local) number operator $\hat{n}_s^i $ as the direct tensor product of the projector $\ket{s}\bra{s}$ at site $i$ and identity matrices on all other sites. The probability of finding node $i$ in state $s$ at time $t$ is then:
\begin{equation}\label{projected}
	P(X_i(t) = s) = \langle \hat{n}_s^i (t) \rangle = \langle \mathbf{1} | \hat{n}_s^i |\Phi(t)\rangle \equiv \bracket{\hat{n}_s^i}{\Phi(t)}\,.
\end{equation} 
Here $\bra{\mathbf{1}}$ denotes the flat state, a column vector with all elements equal to one and we have defined the projection co-vector $\bra{\hat{n}_s^i} =\langle \mathbf{1} | \hat{n}_s^i $. 

Projections are equally well suited to obtain conditional probabilities, where the conditions are implemented as projections on the initial configurations. For instance, the conditional probability of finding node $i$ in state $s_1 \in \Sigma$ at time $t$, given that node $j$ was in state $s_2 \in \Sigma$ at $t=0$ is given by:
\begin{equation}
	P(X_i(t) = s_1 | X_j(0) = s_2) = \frac{ \langle \hat{n}_{s_1}^i | \hat{U}(t) | \hat{n}_{s_2}^j \rangle }{\bracket{\mathbf{1}}{\hat{n}_{s_2}^j}  }
\end{equation}
The normalization in the denominator is there to ensure that $\ket{\Phi(0)} = \ket{\hat{n}_{s_2}^j}/\bracket{\mathbf{1}}{\hat{n}_{s_2}^j}  $ is a normalized probability distribution corresponding to all microscopic configurations in which node $j$ is in state $s_2$. 

After defining the projection operators, the dynamics of node-based observables can be inferred from the configuration space basis, by using the master equation \eqref{mastereqn} inside the projections \eqref{projected}. This gives the following general equation for the time evolution of node-based densities:
\begin{equation}\label{nodebased_evo}
	\partial_t \langle \hat{n}_s^i (t) \rangle = \bra{\hat{n}_s^i } \hat{W}(A(t)) \ket{\Phi(t)} = \bra{\mathbf{1}} [\hat n_s^i, \hat{W}] \ket{\Phi(t)}\,,
\end{equation}
where the last line follows from the fact that $\hat{W}$ is a Markov generator, satisfying $\bra{\mathbf{1}} \hat{W} = 0$. Equation \eqref{nodebased_evo} shows that the time-evolution for the probability of node $i$ to be in state $s$ is generated by the expectation value of the commutator between the local number operator for state $s$ and the Markov generator. 

Similarly, if one is interested in global observables, such as the density of nodes in state $s$, one can straight-forwardly project on sums of local operators. For instance:
\begin{equation}\label{densop}
	\langle \hat{n}_s \rangle = \frac{1}{N} \sum_{i = 1}^N  \bracket{\hat{n}_s^i}{\Phi(t)}\,,
\end{equation}
computes the expected density of nodes in state $s$. The configuration-based approach, however, contains more information than just expected values. In principle, all variances and higher moments are contained within the propagator $\hat{U}(t)$. Particularly, one might be interested in the moment generating function for the density of nodes in state $s$, which depends on a dual variable $\mu$. This is computed as the expectation value of the matrix exponential of $\mu \hat{n}_s$, which gives explicitly:
\begin{align}\label{mgf}
	M_s(\mu) & = \vev{e^{\mu \hat{n}_s} } = \bra{\mathbf{1}} e^{\frac{\mu}{N} \sum_{i = 1}^N \hat{n}_s^i } \ket{\Phi(t)} \\
	& = \bra{\mathbf{1} } \prod_{i=1}^{N} e^{\frac{\mu}{N} \hat{n}_s^i} \ket{\Phi(t)} \equiv \bracket{\mu}{\Phi(t)}\,.
\end{align}
 Here we have defined the co-vector $\bra{\mu} = \bra{\mathbf{1} } \prod_{i=1}^{N} e^{\frac{\mu}{N} \hat{n}_s^i}$, which may be written as the direct product of $N$, $d$-dimensional co-vectors, which have all unit entries, except at state $s$, where it equals $e^{\mu/N}$.

\section{Applications to network epidemiology}\label{sec:applications}
The above, somewhat formal, discussion is made more concrete by examples. In \cite{prldraft} we considered two models with identical node-based dynamics, which nonetheless show wildly different behavior in terms of the configuration space description: the voter model and the simple exclusion process on networks. Here we will consider in more detail the application to models of network epidemiology. In particular, we will look at the susceptible-infected-susceptible (SIS) and susceptible-infected-recovered (SIR) models, as examples of stochastic processes with an out-of-equilibrium (dynamical) phase transition.

\subsection{SIS and SIR models}
To model a spreading process such as the susceptible-infected-susceptible (SIS) model, one should choose the operators in \eqref{genericW} to reflect the appropriate transitions at the individual level. For the SIS model, the two dimensional state space is spanned by the states $\Sigma = \{\S, \I \}$ for susceptible and infected nodes, respectively. By recovery, infected nodes transition to the susceptible state, independent of its neighbors states. Hence, recovery is implemented by a local operator $\hat q_{\I \to \S}$, given as a matrix as:
\begin{equation}
\hat q_{\I \to \S} = \ket{\S}\bra{\I} - \ket{\I}\bra{\I}\,.
\end{equation}
The operator $\hat q_{\I \to \S}^i$ is then the tensor product of $ \hat q_{\I \to \S}$ inserted at node $i$ and $N-1$ identity operators for all other sites. 

The second transition of the SIS model is the transmission of the infection. In this case two neighboring nodes are needed. If any node $i$ is infected, it can infect its susceptible neighbors $j$. This transition is implemented by a bilinear operator $\hat n^i_{\I} \, \hat q^j_{\S \to \I}$ where:
\begin{equation}\label{transop}
\hat n_{\I} = \ket{\I}\bra{\I}\,, \qquad \hat q_{\S \to \I} = \ket{\I}\bra{\S} - \ket{\S} \bra{\S}\,.
\end{equation}
The transition rate matrix for the SIS model on a network $A^{ij}$ is then
\begin{equation}\label{wsis}
\hat W_{\rm SIS}(A) = r_{\rm trans} \sum_{i,j = 1}^N A^{ij} \hat n^i_{\I} \, \hat q^j_{\S \to \I} + r_{\rm rec} \sum_{i = 1}^N \hat q_{\I \to \S}^i \,.
\end{equation}
In the case of time-independent network interactions, the solution to the master equation \eqref{mastereqn} is 
\begin{equation}
\ket{\Phi(t)} = e^{ \hat W_{\rm SIS}(A) t} \ket{X_0} \,.
\end{equation}
where $\ket{X_0}$ denotes the initial configuration of the network. 

For the SIR model, recovered nodes are no longer susceptible, but move to a third compartment $\R$. Hence, in this case the state space is three dimensional ($\Sigma = \{\S, \I, \R\}$) and the local recovery transmission is implemented by the infinitesimal stochastic matrix $\hat q_{\I \to \R} = \ket{\R} \bra{\I} - \ket{\I} \bra{\I}$. As such the Markov generator for the SIR model on networks is
\begin{equation}
	\hat W_{\rm SIR}(A) = r_{\rm trans} \sum_{i,j = 1}^N A^{ij} \hat n^i_{\I} \, \hat q^j_{\S \to \I} + r_{\rm rec} \sum_{i = 1}^N \hat q_{\I \to \R}^i \,.
\end{equation}
Any configuration without infected nodes is an absorbing state in both the SIS and SIR models.

\subsection{Limit to node-based dynamics}

As shown in section \ref{sec:nodebasedobs}, the local node-based observables are recovered by projections over all configurations where the node of interest is fixed in a given state. Here we will show how well-known mean-field approximations for epidemic models on networks are recovered from this formulation. We will illustrate this for the SIS model, the node-based dynamics in the SIR model can be recovered by similar arguments. We will henceforth scale time as $t \to r_{\rm rec}t$ and work only with the dimensionless ratio of transition rates $\lambda = r_{\rm trans}/r_{\rm rec}$. The expectation value for node $i$ to be in the infected state is given as 
\begin{equation}
	\vev{\hat{n}_{\I}^i (t)} = \bracket{\hat{n}_{\I}^i}{\Phi(t)}\,.
\end{equation}
As shown above, its time evolution is given by the expectation value of the commutator of $\hat{n}_{\I}^i$ with the Markov generator $\hat{W}_{\rm SIS}$. This works out to be:
\begin{equation}
	[\hat{n}_\I^i , \hat{W}_{\rm SIS}] = \lambda  \sum_{k} A^{ki} \hat{n}_\I^k \, \hat \sigma_-^i -  \hat \sigma_+^i 
\end{equation}
where we have defined the matrices $\hat \sigma_- = \ket{\I} \bra{\S}$ and $\hat \sigma_+ = \ket{\S} \bra{\I}$. These matrices are off-diagonal, however, due to the contraction with the flat state on the left, their expectation values are identical to those of the number operators $\hat{n}_{\S}$ and $\hat{n}_{\I}$, respectively. Using this in equation \eqref{nodebased_evo} gives the node-based evolution for $\vev{\hat{n}_{\I}^i (t)}$ as
\begin{equation}\label{nodebasedSIS}
	\partial_t \vev{\hat{n}_{\I}^i (t)} = \lambda \sum_{k=1}^{N} A^{ki} \vev{\hat{n}_\I^k \hat{n}_\S^i(t)} - \vev{\hat{n}_{\I}^i (t)}\,,
\end{equation}
where here $ \vev{\hat{n}_\I^k \hat{n}_\S^i(t)} $ is the expectation value for the nodes $i$ and $k$ to form an $\S\I$ pair. The dynamical evolution of the node-based system hence depends on the presence of $\S\I$ pairs, which in turn depends on triples and pairs. Indeed, by working out the commutator $[\hat{n}_\I^k \hat{n}_\S^i, \hat{W}_{\rm SIS}]$ and taking the expectation value, it is not hard to find the dynamical evolution equation for $\S\I$ pairs:
\begin{align}
	\partial_t  \vev{\hat{n}_\I^k \hat{n}_\S^i(t)} & = \lambda \sum_{l=1}^N  \left(  A^{lk} \vev{\hat{n}_{\I}^l \hat{n}_\S^k \hat{n}_\S^i (t)} - A^{li} \vev{ \hat n_{\I}^k \hat n_\S^i \hat{n}_{\I}^l (t)} \right) \nonumber \\
		& + \vev{\hat{n}_\I^k \hat{n}_\I^i(t)} -  \vev{\hat{n}_\I^k \hat{n}_\S^i(t)} \,.
\end{align}

The dependence of single node expectation values on pairs and the generation of a nested hierarchy of coupled differential equations is well-known in network epidemiology (see for instance \cite{kiss2017mathematics}). Generally, the chain of interdependent differential equations is `closed' at some level by a mean-field approximation. At the first level, this can be done by approximating in \eqref{nodebasedSIS} the expectation value for $\S\I$ pairs as a direct product of single-node expectation values: $\vev{\hat{n}_\I^k \hat{n}_\S^i(t)} \sim \vev{\hat{n}_\I^k(t)} \vev{\hat{n}_\S^i(t)}$. Under this mean-field--type assumption, equation \eqref{nodebasedSIS} becomes the $N$-intertwined SIS epidemic network model of \cite{van2008virus,van2011n}. Closures at higher levels are also possible, by approximating expectation values for triples as a combination of pair and single node expectation values \cite{kiss2017mathematics}. 
However, in principle the complete and exact hierarchy of coupled differential equations can be derived from the configuration based information dynamics as outlined in this section.

\begin{figure*}[!ht]
	\centering
	\includegraphics[width=0.9\linewidth]{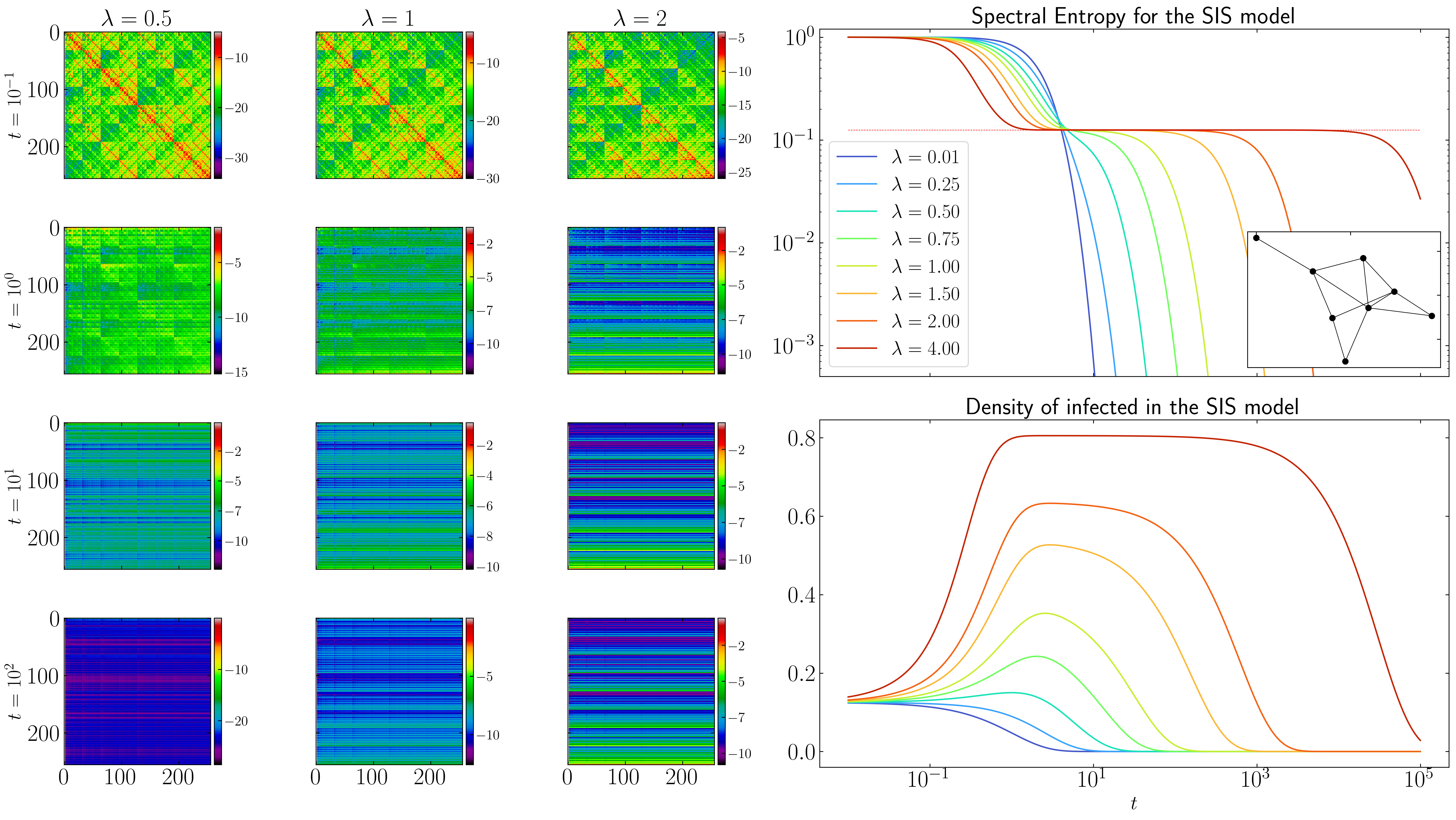} 
	\caption{\label{fig:SvNsis} Left: heat plots of the logarithm of the density matrix $\hat{\rho}(t)$ for the SIS model at different times and effective infection rate $\lambda = r_{\rm trans}/r_{\rm rec}$. The columns of the plot label the initial configuration $X$ (as an integer whose binary representation corresponds to the network configuration), while the rows label the microstates $Y$ at time $t$ indicated on the left hand side. As such, each pixel represent the information flow from configuration $X$ (column) to the configuration $Y$ (row). Right: the (normalized) von Neumann entropy $\mathcal{S}(t)$ for the SIS model at several values of $\lambda$ (top) and the density of infected nodes in the network starting from a initial uniform distribution over configurations with a single infected node (bottom). The computations were performed using the exact transition rate matrix on the small eight-node network displayed in the inlay.} 
\end{figure*}

\section{Exact diagonalization}\label{sec:exactdiag}

For small networks ($N \sim 10$), the transition rate matrices can be worked out explicitly using exact diagonalization of the exponential of the Markov generator $\hat{W}$. In this section we will show results for the applications relating to network epidemiology for low-dimensional networks where exact diagonalization is still possible. Note that our aim is to provide an exemplary application of our framework: more realistic settings dealing with larger systems require more sophisticated analytical tools to deal with the underlying computational complexity. In the concluding section \ref{sec:conclusion}, we describe how this work can be extended to deal with these more challenging scenarios.  

\begin{figure*}[!ht]
	\centering
	\includegraphics[width=0.8\linewidth]{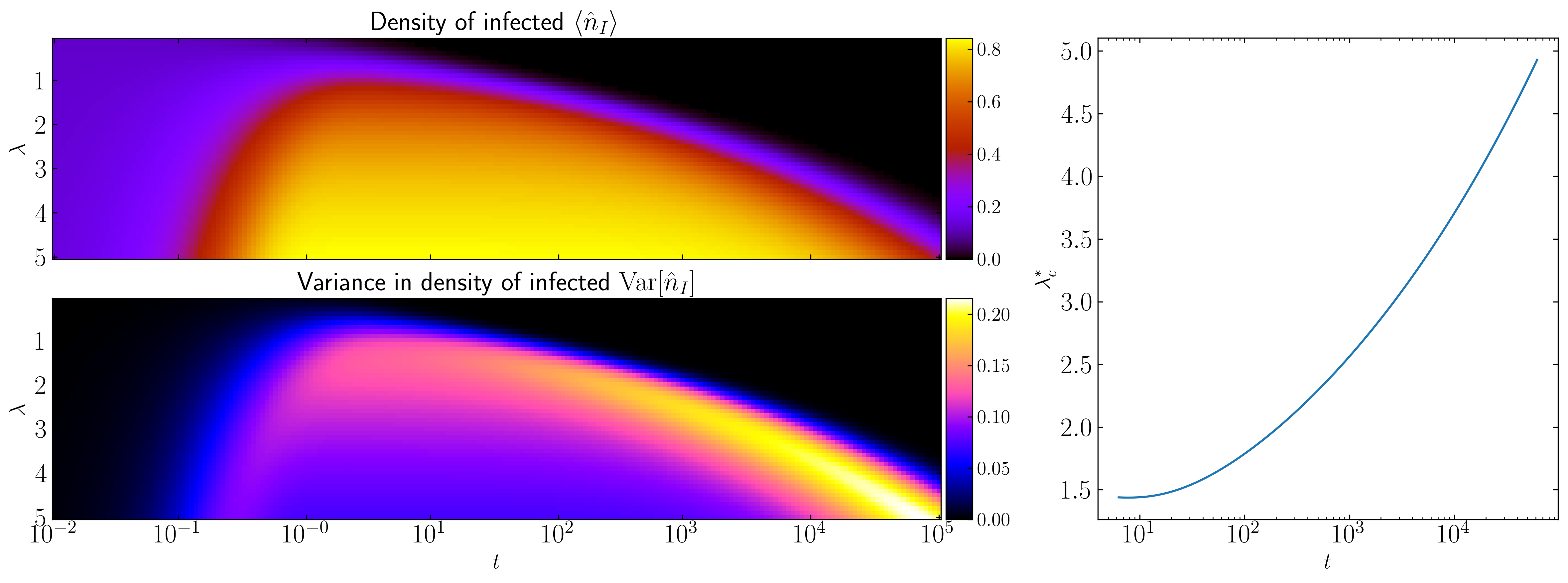} 
	\caption{\label{fig:lambdac} Left top: The density of infected at various $\lambda$ as a function of time. The initial conditions for this plot were taken to be the uniform distribution over all configurations with a single infected node (and all other susceptible). Left bottom: The variance in density of infected nodes for various $\lambda$ as a function of time.
	Right: The variance of the density of infected peaks at a certain $\lambda_c^*(t)$, depending on the time-scale $t$ at which one observes the system. This finite-size critical value $\lambda_c^*$ goes to infinity as time progresses.  
	} 
\end{figure*}

\subsection{SIS model}

In figure \ref{fig:SvNsis} we display the density matrices and von Neumann entropy for the SIS model on a small network at several values for the ratio $\lambda = \frac{r_{\rm trans}}{r_{\rm rec}}$. We see that for low values of $\lambda$, the entropy immediately decays from the maximal value of $N \log(2)$ to 0, reflecting a fast relaxation to the healthy absorbing state. For large values of $\lambda$, the entropy seems to reach a non-zero value, however, it continues to decay at a much slower rate. Eventually, even for large values of $\lambda$, the entropy will vanish at late times, due to the fact that the Markov chain has a single absorbing state: the healthy population. Even if the infection rate is large, stochastic fluctuations will ensure that the absorbing state is reached eventually, at which point the dynamics will stop.  In this sense, there is no critical threshold for finite sized systems, as the late time behaviour is always the absorbing state. The critical threshold $\lambda_c$ is only well-defined in the limit of infinite system size, or in the mean-field approximation.  

%The formation of a region with non-zero spectral entropy at intermediate times is a signal of the endemic phase transition in the SIS model: for $\lambda > \lambda_c$, a meta-stable plateau in $\mathcal{S}(t)$ emerges, indicating the presence of trapped field other than the absorbing state. 

\begin{figure*}[!ht]
	\centering
	\includegraphics[width=0.9\linewidth]{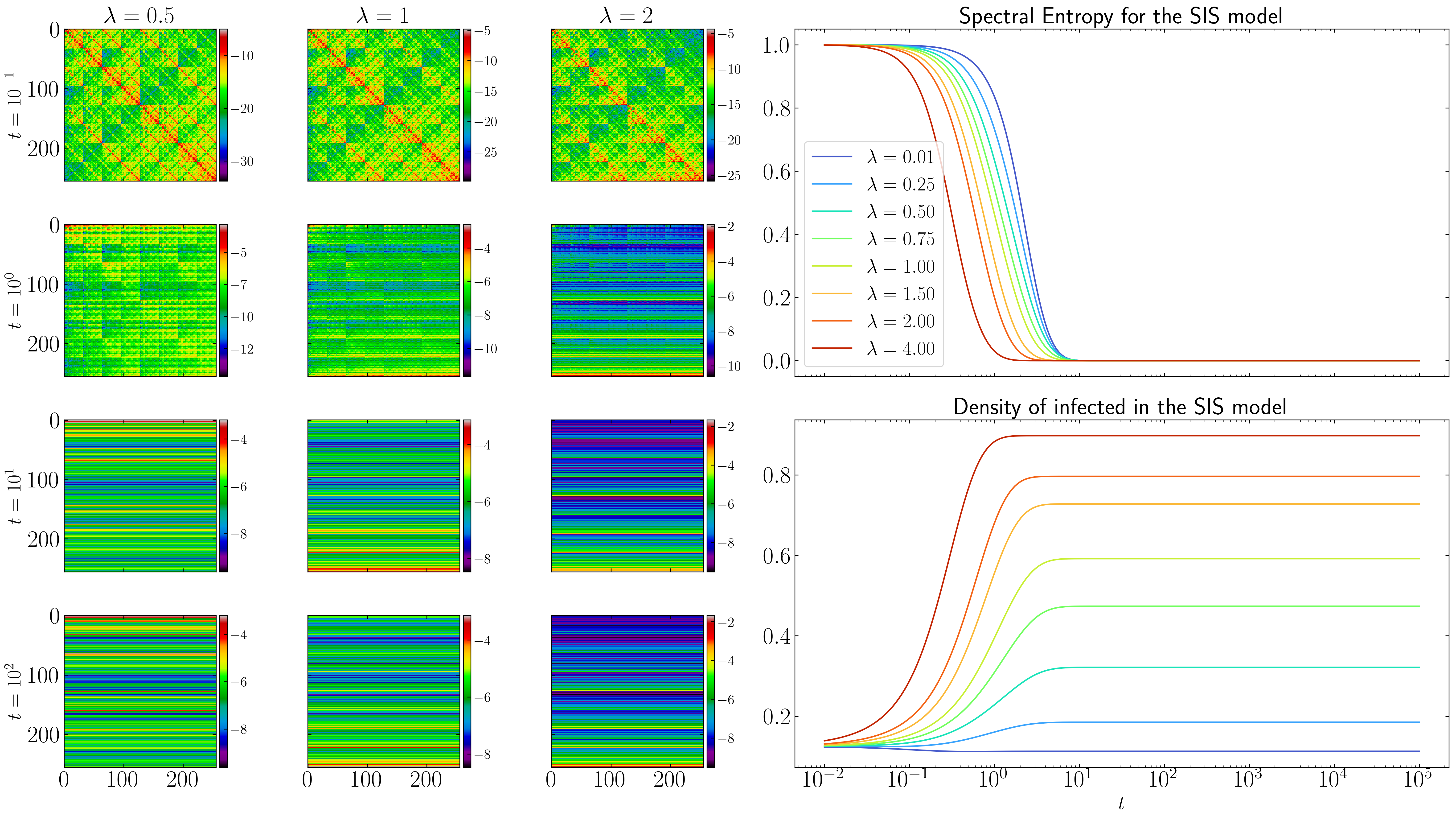} 
	\caption{\label{fig:SvNsis_driven} Similar plot as Fig. \ref{fig:SvNsis} now with the inclusion of the driving term \eqref{driving} (with $\omega = 1$). The absorbing (healthy) state is now no longer an absorbing state, which removes the plateau in the spectral entropy for $\lambda > \lambda_c$. The driven system has a single non-equilibrium steady state to which the dynamics flows, regardless of initial conditions. The density of infected nodes now reaches a finite value for all values of $\lambda$.} 
\end{figure*}

However, it is still possible to find signatures of the endemic state with non-zero density of infected nodes. For large $\lambda$, a meta-stable plateau in $\mathcal{S}(t)$ emerges, which shows the presence of trapped field other than the absorbing state. The value of the entropy at the plateau is $\log(2)$, indicating the presence of two simultaneously activated information channels, one being the absorbing state and the other an endemic state where the system has non-zero prevalence. This can be seen by comparing with the density of infected $\vev{\hat{n}_\I}$ in the bottom panel on the right of Figure \ref{fig:SvNsis}. The density was computed using initial conditions where a single source node is infected at time $t=0$, averaged over all initial source nodes. As such, the initial state $\ket{\Phi(0)}$ is the uniform distribution over all network configurations with a single infected node.

The absence of a critical threshold $\lambda_c$ in the finite size system is also illustrated in Figure \ref{fig:lambdac}. There we plot the expected density of infected nodes and its variance as a function of $\lambda$ and time. The variance in density shows a peak at a certain finite size (and finite time) critical point $\lambda_c^*(t)$, which is associated with the onset of criticality. The right panel of Fig.~\ref{fig:lambdac} indicates that the location of the peak diverges at $t\to \infty$, hence the critical threshold disappears in the infinite time limit.  

The absorbing state for the SIS model can be removed by including a `driving term' into the dynamics. By infecting a random node whenever the absorbing state is reached, the system can be driven to a non-equilibrium steady state (NESS). In our framework, this can be implemented by the addition the following driving term to the Markov generation \eqref{wsis}: 
%The presence of the absorbing state complicates the stochastic thermodynamical interpretation of the finite size SIS model. \textcolor{blue}{This is due to the fact that for any finite size network, the healthy absorbing state is always reached, even above the critical threshold, due to random stochastic fluctuations.} 
%To overcome this, the system can be driven to a non-equilibrium steady state (NESS), by randomly infecting a node when the absorbing state is reached. In our framework, this can be easily implemented by adding a `driving term' to the Markov generator. There are many possible ways to do so, here we add the term:
\begin{equation}\label{driving}
    \hat{W}_{\rm driv.} = \omega \sum_{i=1}^N  \hat{n}_{\I}^i  \prod_{ \substack{j = 1 \\ j \neq i}}^N  \hat{n}_{\S}^j  \,.
\end{equation}
%where $\hat{n}_{\S} = \ket{\S}\bra{\S}$ and $\hat{b}_{\rm trans}$ is defined in \eqref{transop}. 
This term effectively removes the absorbing state by randomly infecting a single node (with rate $\omega$) whenever the healthy state is reached. Interestingly, as shown in Figure \ref{fig:SvNsis_driven}, the presence of a driving term removes the previously seen plateau in the spectral entropy, which now drops to zero immediately for all values of the transmission rate $\lambda$, while the density remains finite at late times. This indicates that there is now a single information channel at late times, corresponding to the NESS of the driven system. The lower right panel of Fig.~\ref{fig:SvNsis_driven} shows that the NESS contains a non-trivial density of infected for all values of $\lambda$, which increases with $\lambda$.

\begin{figure}[!t]
	\centering
	\includegraphics[width=\linewidth]{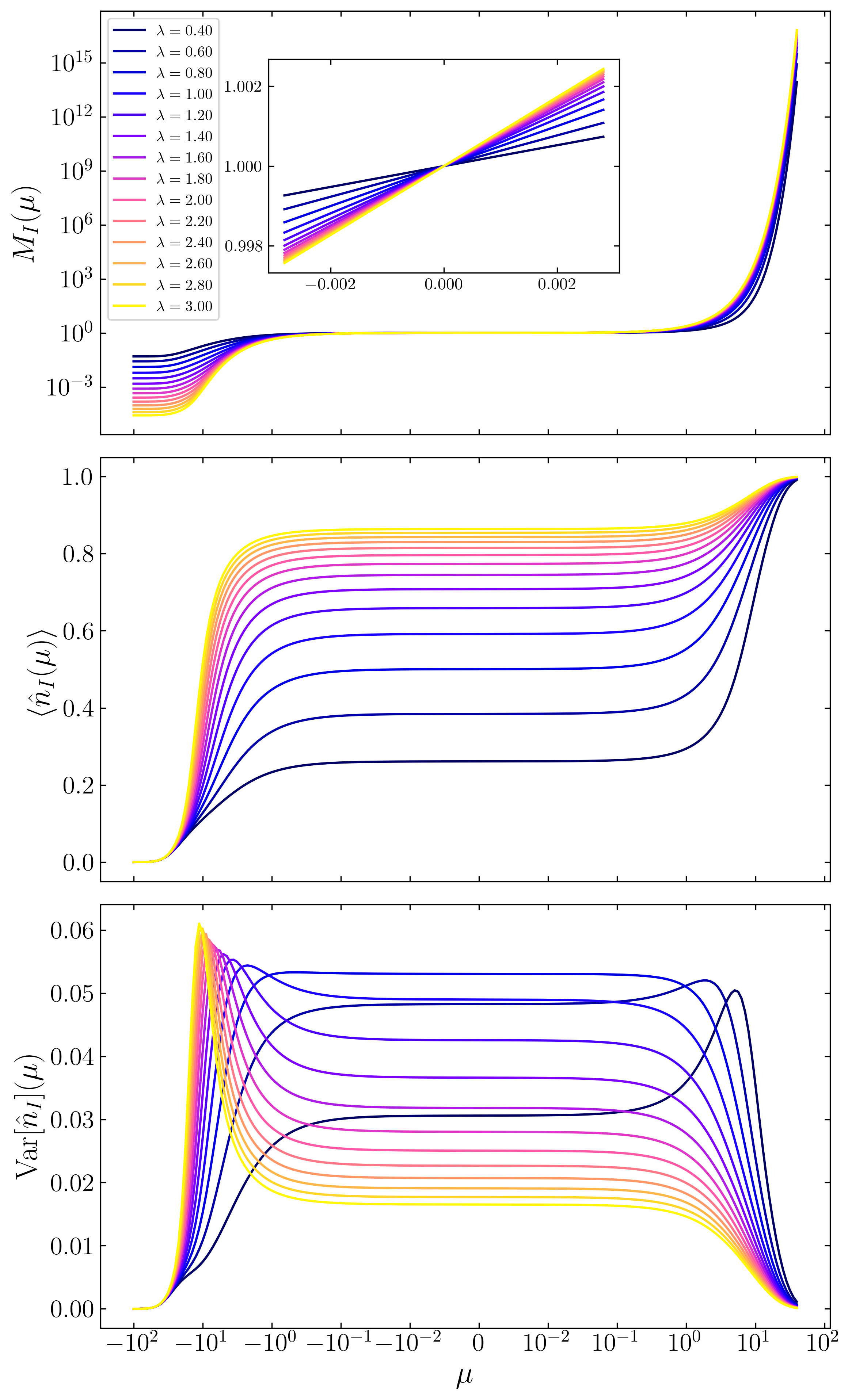} 
	\caption{\label{fig:mgfsis} Top: The moment generating function in the NESS of the driven SIS model on the 8-node network of Figure \ref{fig:SvNsis}. Middle: Expectation value of the density of infected nodes as a function of dual parameter $\mu$. Bottom: Variance in the density of infected nodes as a function of $\mu$.} 
\end{figure}

Now that we have obtained the configuration space description for the NESS of the driven system, we can compute the moment generating function $M_I(\mu)$ for the global density of nodes in the infected state using equation \eqref{mgf}. We plot this moment generating function for several $\lambda$ values in the top panel of Figure \ref{fig:mgfsis}. The expected density of infected nodes can be obtained by derivation of $\log(M_I(\mu))$ with respect to $\mu$, followed by setting $\mu$ to zero. The middle panel of Figure \ref{fig:mgfsis} shows $ \vev{\hat{n}_\I (\mu)} = \partial_\mu \log(M_I(\mu))$, illustrating not only the expected density at $\mu=0$, but also the behavior of the function at positive and negative $\mu$ values. These correspond to ensemble averages over trajectories tilted towards more or less dense configurations, respectively. 

The bottom panel of Figure \ref{fig:mgfsis} shows the second derivative $\partial_\mu^2 \log(M_i(\mu))$, which at $\mu=0$ corresponds to the variance in density of infected nodes. One can observe that the peak in variance crosses the $\mu=0$ line as $\lambda $ is decreased. This implies that for a certain value of $\lambda$, the variance in density peaks at $\mu =0$, which corresponds to the finite size analogue of a critical transmission rate $\lambda_c$.

\subsection{SIR model}

\begin{figure}[!ht]
	\centering
	\includegraphics[width=\linewidth]{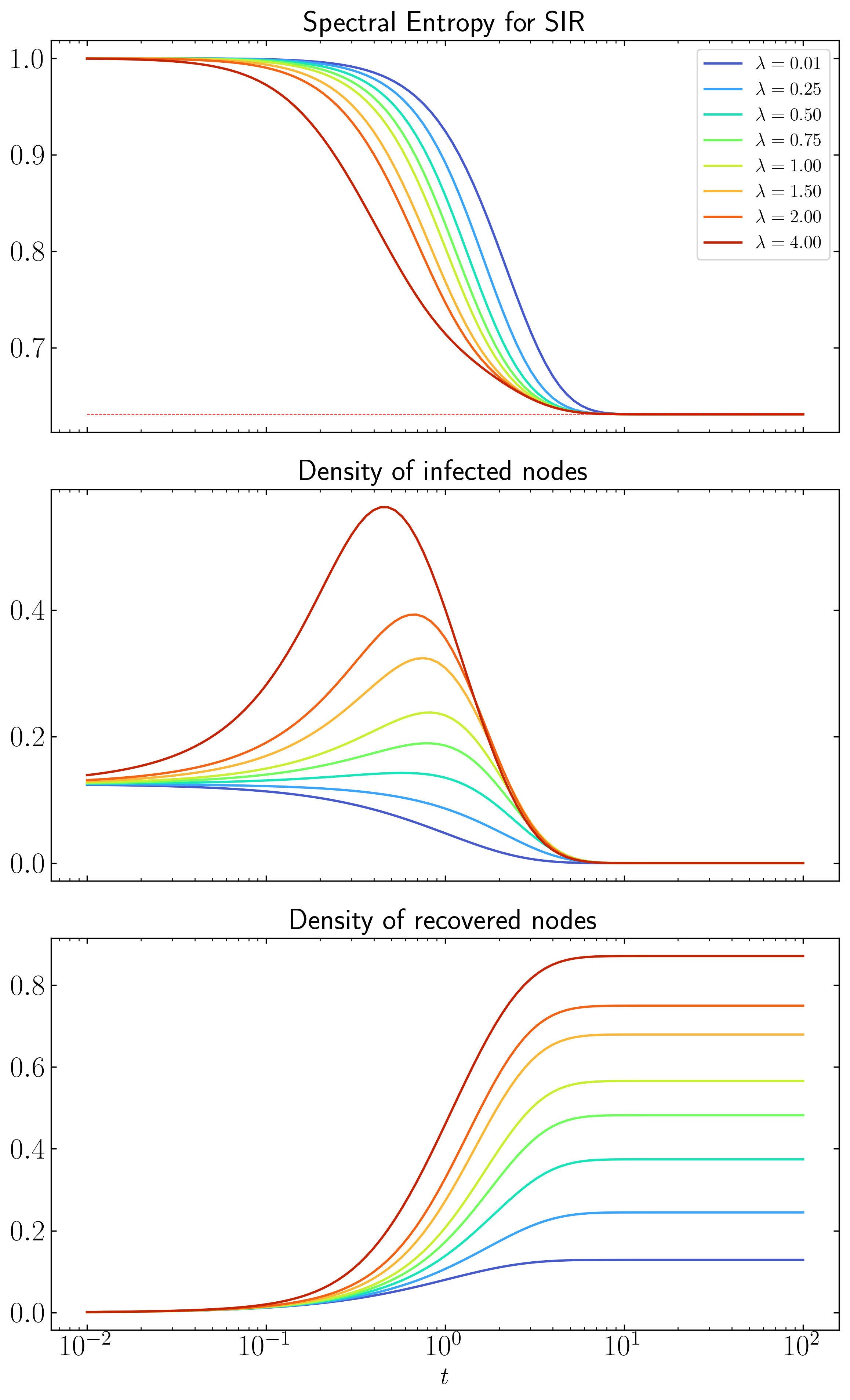} 
	\caption{\label{fig:sir_SE} Top: the (normalized) spectral entropy of the SIR model on the 8-node network shown in Figure \ref{fig:SvNsis}. Middle: the density of infected nodes in the SIR model on the same network, starting from an initial uniform distribution over configurations with only a single infected node. Bottom: the density of recovered nodes as a function of time, using the same initial conditions} 
\end{figure}

The top panel of Figure \ref{fig:sir_SE} shows the spectral entropy for the SIR model on the same graph as shown in the inlay of Figure \ref{fig:SvNsis}, normalized by dividing $\mathcal{S}(t)$ by its maximal value $N \log(3)$. In this case, any state without infected nodes is absorbing, hence the spectral entropy relaxes to $N \log(2)$ at late times, illustrated by the dotted red line. The middle panel shows the expected density of infected nodes in the network, obtained by using equation \eqref{densop} for $\hat{n}_{\I}$. The initial conditions of this plot were again chosen as the uniform distribution over all configurations where only a single node is infected, while all other nodes are in the susceptible state. The bottom panel shows the density in the recovered compartment, using the same initial conditions.

Using the mathematical techniques described here, extensions to more complex spreading models, such as the $\S\I\R\S$ or $\S\textsc{e}\I\R$ models, are straightforward, although the analysis using exact diagonalization will fail rapidly for larger networks. In these cases, however, we expect the exponential complexity of the configuration space description to be highly redundant. In fact, it could very well be that for larger systems governed by local dynamics on sparse networks, the exact description can be compressed to a lower dimensional subspace, growing only polynomially in system size. It would be very interesting to investigate under which circumstances such an efficient compression can take place. We comment in the conclusions below on one possible avenue for future research along this direction.

%{\bf WM: what more to say? Extend to SIRS, possibly? This will remove most of the absorbing states (except for the healthy state...). Might take a very long time to run, though} {\bf MDD: nope. I would instead add a short paragraph on how to extend this work to other compartmental models, outlining why it's difficult and not done here, and how this can be solved with more sophisticated techniques for many-body analysis}

\section{Conclusion}\label{sec:conclusion}

In this paper and the companion letter \cite{prldraft}, we have extended the statistical field theory of complex information dynamics to a configuration based representation, suitable for models where nodes represent vector-valued information fields. 
This is relevant for a broad class of models, most notably those governed by pair-wise mediated Markovian dynamics. Here, we have illustrated the general theory at the hand of examples from network epidemiology. Note that the analytical technique used for the analysis can be employed only for low-dimensional systems: since this is the first practical example of this new framework, we have opted to provide the general reader with an exemplary application. We have discussed how to recover the node-based information dynamics from suitable projections on sub-spaces of the configuration based dynamics. In the example of epidemic models this recovers the hierarchy of coupled differential equations describing the exact node-based dynamics of the model. We have also computed the spectral entropy by exact diagonalization for different models of network epidemiology on a small sample network,  which demonstrates that it is a measure of the diversity of available information channels on the space of network configurations. 

The high dimensionality of the configuration space makes the propagator $\hat{U}(t)$ difficult to compute explicitly for large networks. In this work, we have focused on exact diagonalization, feasible only for small networks, hence it is desirable to explore other methodologies suitable for larger networks. In quantum many-body systems a similar problem occurs, where wavefunctions of composite quantum systems are vectors in the tensor product Hilbert space of exponentially large complexity. However, for quantum systems with local interactions, the physically relevant Hilbert space is in fact much smaller \cite{poulin2011quantum}. State of the art methods in the numerical approximation of composite quantum systems use this feature to find efficient representations of exponentially large vectors using tensor networks  \cite{white1992density,schollwock2011density,orus2014practical}. These methods can similarly be applied to stochastic systems with many interacting constituents and subject to Markovian dynamics. Hence, a natural extension of this work is to explore how tensor networks can be used to provide efficient and accurate solutions of the configuration space complex information dynamics described in this work. Recent work using tensor networks for studying large deviation statistics in lattice models of stochastic systems are  \cite{helms2019dynamical,banuls2019using,causer2021optimal,strand2022using,garrahan2022topological,causer2022finite}. It would be very interesting to extend this research direction by investigating which types of network structures and dynamics allow for a efficient compression of the exact description, while retaining accurate information on all relevant observables in the system.

\begin{acknowledgments}
MDD acknowledges financial support from the Human Frontier Science Program Organization (HFSP Ref. RGY0064/2022), from the University of Padua (PRD-BIRD 2022) and from the EU funding within the MUR PNRR “National Center for HPC, BIG DATA AND QUANTUM COMPUTING” (Project no. CN00000013 CN1).
\end{acknowledgments}

%\bibliography{biblio}


\end{document}